\documentclass[notitlepage,onecolumn,tighten,longauthor,12pt]{revtex4-1}

\usepackage{amssymb,amsfonts,amsmath}
\usepackage{subfigure}
\usepackage{rotating}
\usepackage{graphicx}
\usepackage{bm}
\usepackage{hyperref}
\usepackage{url}
\usepackage{epstopdf}
\usepackage{multirow}
\usepackage{enumerate}
\usepackage{natbib}

\newcommand*{\bibnamefont}[1]{#1}
\newcommand*{\bibfnamefont}[1]{#1}

\ifx\pdfoutput\undefined
\usepackage[dvips,bookmarks]{hyperref}	
\else
\fi

\usepackage{color}
\definecolor{rosso}{cmyk}{0,1,1,0.4}
\definecolor{rossos}{cmyk}{0,1,1,0.55}
\definecolor{rossoc}{cmyk}{0,1,1,0.2}
\definecolor{blu}{cmyk}{1,1,0,0.3}
\definecolor{blus}{cmyk}{1,1,0,0.6}
\definecolor{bluc}{cmyk}{1,1,0,0.1}
\definecolor{verde}{cmyk}{0.92,0,0.59,0.25}
\definecolor{verdec}{cmyk}{0.92,0,0.59,0.15}
\definecolor{verdes}{cmyk}{0.92,0,0.59,0.4}
\definecolor{bviolet}{rgb}{0.54, 0.17, 0.89}
\hypersetup{colorlinks,bookmarksopen,bookmarksnumbered,citecolor=bviolet,
linkcolor=bviolet,pdfstartview=FitH,urlcolor=bviolet}

\begin{document}

\title{Dark matter self-interactions from the internal dynamics of dwarf spheroidals}
\author{Mauro Valli$^{1}$}
\author{Hai-Bo Yu$^{2}$}

\maketitle

\textbf{Dwarf spheroidal galaxies provide well-known challenges to the standard cold and collisionless dark matter scenario \citep{Tulin17,Bullock17}: The \textit{too-big-to-fail} problem, namely the mismatch between the observed mass enclosed within the half-light radius of dwarf spheroidals \citep{Walker09,Wolf10} and cold dark matter N-body predictions \citep{Boylan11,Boylan12};  The hints for inner constant-density cores \citep{Battaglia08,Walker11,Amorisco12,Strigari17}. While these controversies may be alleviated by baryonic physics and environmental effects \citep{Pontzen14,Sawala16,Dutton16,Wetzel16,Fattahi16}, 
revisiting the standard lore of cold and collisionless dark matter remains an intriguing possibility.
Self-interacting dark matter \cite{Spergel00,Kaplinghat16} may be the successful proposal to such a small-scale crisis \citep{Vogelsberger12,Rocha12}. Self-interactions correlate dark matter and baryon distributions, allowing for constant-density cores in low surface brightness galaxies \citep{Kaplinghat14,Elbert15,Kamada17,Creasey17}.
Here we report the first data-driven study of the \textit{too-big-to-fail} of Milky Way dwarf spheroidals within the self-interacting dark matter paradigm. We find good description of stellar kinematics and compatibility with the concentration-mass relation from the pure cold dark matter simulation in~\cite{Vogelsberger16}. Within the latter, a subset of Milky Way dwarfs are well fitted by cross sections of 0.5--3 cm$^2$/g, while others point to values greater than 10~cm$^2$/g.}

~

The internal dynamics of dwarf spheroidal galaxies (dSphs) of the Milky Way (MW) is commonly studied exploiting the kinematics of a stellar population of density $\rho_{\star}$, in dynamical equilibrium under the gravitational potential governed by dark matter (DM), $ \Phi_{\textrm{\tiny DM}}$. For a spherically symmetric steady-state system, the first moment of the collisionless Boltzmann equation for the stellar phase-space distribution takes the form:
\begin{equation}
\left(\rho_{\star} \sigma_r^{2}\right)' + 2 \beta(r)\rho_{\star}(r) \sigma_r^{2}(r)= - \rho_{\star}(r) \, \Phi_{\textrm{\tiny DM}}' \ ,
\label{eq:sph_jeans_eq}
\end{equation}
where the prime denotes logarithmic derivative in $r$. The stellar orbital anisotropy, $\beta \equiv 1-\sigma^{2}_{t}/\sigma^{2}_{r}$, measures the deviation from isotropy in the stellar velocity dispersion tensor.
Photometric observations of the surface brightness of these systems constrain $\rho_{\star}$. Supplying equation~(\ref{eq:sph_jeans_eq}) with Poisson's equation, mass $M(r)$ and orbital anisotropy $\beta(r)$ are inferred from line-of-sight projected spectroscopic measurements, see e.g. \citep{Battaglia13}, leading in general to a degeneracy problem.

The self-interacting dark matter (SIDM) paradigm can be investigated with equation~(\ref{eq:sph_jeans_eq}) by means of a semi-analytical halo model based on simple physical grounds.
Introducing the DM self-scattering rate as $\Gamma \equiv \langle \sigma v \rangle \rho / m$, with velocity averaged cross section, $\langle \sigma v \rangle$, density $\rho$, and particle mass $m$, we can define the scale $r_{1}$ at which $\Gamma \big|_{r_{1}}  \simeq t_{\textrm{age}}^{-1}$, with $ t_{\textrm{age}}$ the typical age of the system. As illustrated in \cite{Kaplinghat14,Kaplinghat16},  $r_{1}$ sets a transition from the regime of an isothermal gas with pressure $p \propto \rho$ ($r \lesssim r_{1}$) to the one of a non-interacting particle ensemble ($r \gtrsim r_{1}$).
For DM dominated systems as MW dSphs, in the inner halo region the SIDM profile is given by:
\begin{equation}
x \,  \ddot{h}+2 \dot{h}  = - x \exp\left(h(x)\right) \, , \, \lim_{x \to 0} h(x) = 0 \, , \, \lim_{x \to 0} \dot{h} = 0 \  ,
\label{eq:Jeans_dSph_SIDM}
\end{equation}
where upper-dots denote derivatives in $x$, where we introduced $x \equiv \sqrt{4 \pi G_{N}\rho_{0}} \, r /\sigma_{0}\,$ with $G_{N}$ the gravitational constant, and $h \equiv \ln \left(\rho / \rho_{0} \right) $. The physical halo from equation~(\ref{eq:Jeans_dSph_SIDM}) is an isothermal cored profile involving the one-dimensional DM velocity dispersion, $\sigma_{0}$, defining the SIDM isothermal gas law, and the central density, $\rho_{0}$.
The solution of equation~(\ref{eq:Jeans_dSph_SIDM}) is then matched at  $r_{1}$ to the Navarro-Frenk-White (NFW) model \citep{Navarro97}.  
The condition of continuity of mass and density at $r_{1}$ implies:
\begin{equation}
\left(\frac{1+x_{s 1}}{x_{s 1}}\right)^{2} \left( \log(1+x_{s 1})-\frac{x_{s 1}}{1+x_{s 1}}\right)   =  \mathcal{R}\big|_{r_{1}} \ ,
\label{eq:matching_CDM}
\end{equation}
with $x_{s 1} \equiv r_{s} / r_{1}$, $r_{s}$ denoting NFW scale radius, and $\mathcal{R}$ being the SIDM ratio $M(r)/\left(4 \pi r^{3} \rho(r)\right)$. Equation~(\ref{eq:matching_CDM}) gives a viable matching if $\mathcal{R}\big|_{r_{1}}>0.5$, which ensures $x_{s 1}>0$. The NFW normalization can be simply read as $ \rho_{s}  =  x_{s 1} (1+x_{s 1})^{2} \rho(r_{1})$.

At given $r_{1}$, we can also estimate  the self-scattering cross section per unit mass of DM particles. This is obtained exploiting the condition $\Gamma \big|_{r_{1}}  \simeq t_{\textrm{age}}^{-1}$, yielding:
\begin{equation}
\sigma/m \simeq \sqrt{\pi} / \left(4 \sigma_{0} \rho(r_{1}) t_{\textrm{age}}\right) \ ,
\label{eq:SIDM_xsec}
\end{equation}
where  $\sigma \simeq \langle \sigma v \rangle /  \langle v \rangle$ and $\langle v \rangle $ is expected to follow a Maxwellian distribution according to the thermalized inner halo by DM self-interactions within $r_{1}$ \cite{Kaplinghat14,Kaplinghat16}.

Using equation~(\ref{eq:sph_jeans_eq}) and such semi-analytical halo model \cite{Kaplinghat14,Kaplinghat16} -- well-tested against SIDM simulations -- we undertake a minimal approach to SIDM paradigm and explore its predictions for the stellar kinematics in MW dSphs. We focus on the eight brightest MW satellites -- the classical dSphs -- regarded as relaxed systems, with small ellipticities \citep{Battaglia13}. They represent the baseline of the too-big-to-fail (TBTF) problem~\cite{Boylan11,Boylan12}. 
We perform fits to the spectroscopic dataset of classical dSphs, adopting a standard Gaussian test statistic. Details about our fitting procedure are given in \textbf{Methods}. 

Main goal of our study is to assess to which extent  TBTF in MW dSphs can be ameliorated within the SIDM scenario. Then, together with the observational information from the stellar kinematics, we also take into account the predictions of pure cold dark matter (CDM) simulations. We specifically exploit here the recent outcome obtained in the N-body study of \cite{Vogelsberger16}, from where we extract the concentration-mass relation:
\begin{equation}
\log_{10} \left(\frac{R_{\textrm{max}}}{\textrm{kpc}}\right) = 0.48 + 0.18 \, \log_{10} \left(\frac{V_{\textrm{max}}}{\textrm{km/s}}\right) \ ,
\label{eq:CDM_mass_conc_rel}
\end{equation} 
within a NFW model with maximum circular velocity $V_{\textrm{max}}$ and corresponding scale radius R$_{\textrm{max}}$. In the whole work, we assume equation~(\ref{eq:CDM_mass_conc_rel}) to be a reliable proxy for the CDM-only  gravitational potential in MW dSphs, assigning $0.2$ dex scatter on $R_{\textrm{max}}$ for $V_{\textrm{max}} \sim 25$--55 km/s, see \textbf{Methods} for further details.

To start with, we report in figure~\ref{fig:SIDM_kinematics_fit} the best-fit result obtained considering only the stellar kinematic data of MW dSphs,  i.e. without imposing the constraint appearing in equation~(\ref{eq:CDM_mass_conc_rel}). We model the stellar orbital anisotropy as constant, $\beta(r) = \beta_{c}$, and consider a cuspy DM halo (orange dotted line) and a cored one (green dotted line), respectively described by NFW and Burkert \cite{Burkert95} profiles. Both cases provide a good fit of dSph stellar kinematics.
This result shows that both cuspy and cored DM halos can yield \textit{a priori} an optimal description of the  kinematic dataset here considered in virtue of the mass-anisotropy degeneracy pertaining to equation~(\ref{eq:sph_jeans_eq}). 

However, if we now focus on the case of cuspy halo profiles and introduce in the fit the constraint from equation~(\ref{eq:CDM_mass_conc_rel}), i.e. we restrict to the representative outcome of pure CDM simulations as in \cite{Vogelsberger16}, an overall satisfactory description of dSph spectroscopic data is no longer available. 
In figure~\ref{fig:SIDM_kinematics_fit} the dot-dashed black line captures this failure representing the NFW scenario subject to the concentration-mass relation with relative scatter as extracted from \cite{Vogelsberger16}.
Most importantly, this fit is performed including also  a more general and realistic orbital anisotropy modeling, namely \cite{Baes07}: 
\begin{equation}
\beta(r) = (\beta_{0}+\beta_{\infty} (r/r_{\beta})^{\eta}) / (1+(r/r_{\beta})^{\eta}) \ , 
\label{eq:beta_r}
\end{equation}
i.e. a spatial interpolation of the regime of stellar motion at the centre, controlled by $\beta_{0}$, and towards the outer region, regulated by $\beta_{\infty}$, with characteristic scale $r_{\beta}$ and slope $\eta$. The data overshooting in several analyzed objects constitutes the essence of the TBTF problem emerging within the CDM picture.

Finally, in the same figure we present the best-fit results for the SIDM scenario. Blue curves correspond to the case where we perform the SIDM fit varying a total of seven parameters, implementing equation~(\ref{eq:beta_r}) together with the 3-parameter semi-analytical halo model, while the dashed light-blue one represents the same SIDM halo model assuming constant stellar anisotropy.
We observe that radial dependence in the stellar orbital anisotropy profile is needed in order to find an overall good fit of dSph kinematics: The general trend for the best-fit anisotropy corresponds to a sharp ($\eta \gtrsim 5$) transition from $\beta_{0} \simeq 0$ to  $\beta_{\infty} \simeq 1$ at $r_{\beta}$, close to the stellar half-light radius. The only exception to this behaviour is provided by Sextans, where circular-like orbits are preferred in the outer region as a consequence of requiring  $r_{1} \lesssim$ 30 kpc (see \textbf{Methods}). 
The underlying NFW profile follows the $R_{\textrm{max}}$~--~$V_{\textrm{max}}$ scatter in the concentration-mass relation  given in equation~(\ref{eq:CDM_mass_conc_rel}), obtained from \cite{Vogelsberger16}.
CDM-only  predictions are recovered above $r_{1}$, highlighted in figure~\ref{fig:SIDM_kinematics_fit}. Therefore, the SIDM paradigm seems to offer a viable proposal for the solution of the TBTF puzzle.

The 7-parameter SIDM fit can be compared against the 6-parameter NFW one -- both subject to the constraint from \cite{Vogelsberger16} --  by means of the Akaike Information Criterion (A.I.C.), whose definition is recalled in \textbf{Methods}. In  figure~\ref{fig:SIDM_kinematics_fit} we report differences $ \Delta \textrm{A.I.C.} = \textrm{A.I.C.}_{\textrm{\tiny{SIDM}}}-\textrm{A.I.C.}_{\textrm{\tiny{CDM}}}$:  Model selection follows according to the smallest possible A.I.C. value. Large differences $ \Delta \textrm{A.I.C.} < 0$ are found for four of the eight MW classicals, showing that stellar kinematics and the N-body input of pure CDM simulations combined together yield a net preference for the SIDM paradigm.

In light of the promising outcome for SIDM resulting from the inspection of the best-fit cases in figure~\ref{fig:SIDM_kinematics_fit}, we continue our study performing a Bayesian analysis of the 7-parameter SIDM model in order to carefully assess the uncertainties on the estimated model parameters through the evaluation of their posterior probability density function (p.d.f.). Following equation~(\ref{eq:SIDM_xsec}), we eventually aim to marginalize over nuisance parameters in order to derive the first data-driven estimate in literature of the self-scattering cross section of DM particles in MW dSphs. Details on the Markov chain Monte Carlo (MCMC) analysis performed for the Bayesian fit of the SIDM model are reported in \textbf{Methods}.

In table~\ref{tab:SIDM_dSph_fit} we collect for each dSph the mode and the $68\%$ highest posterior density (h.p.d.) interval of the marginalized distribution for the fitted parameters. 
The SIDM best-fit in figure~\ref{fig:SIDM_kinematics_fit} approximately lies within $68\%$ h.p.d. of parameters posterior. From the estimated values for $r_{\beta}$, we highlight Sculptor, Sextans and Fornax to be the most representative cases of relevance for spatially dependence in  the stellar orbital anisotropy profile.

Using the MCMC events for the estimated SIDM parameters $\sigma_{0}$, $\rho_{0}$ and $r_{1}$ and marginalizing over orbital anisotropy ones, we can derive the posterior p.d.f. for $r_{s}$ and $\rho_{s}$ of the matched NFW profile at $r_{1}$ by means of equation~(\ref{eq:matching_CDM}). Eventually, we compute the posterior distribution for the SIDM circular velocity profile $V_{c}(r) = \sqrt{G_{N} M(r)/r}$ at a given radius $r$.
In figure~\ref{fig:SIDM_TBTF}, we show the typical spread on the SIDM circular velocity profile for each dSph within the 10-$th$ and 90-$th$ percentile of $V_{c}$ posterior p.d.f. at different radii.
In the outer region, we highlight the expected agreement with the CDM-only output from \cite{Vogelsberger16}, represented by the gray band. In the inner region, we find satisfactory match to dSph half-light masses originally proposed in \cite{Wolf10}: Colored squares represent nominal values for $V_{c}(r_{1/2}) \simeq \sqrt{3 \, \langle \sigma_{los}^{2} \rangle}$, here re-estimated as detailed in \textbf{Methods}.
Figure~\ref{fig:SIDM_TBTF} reinforces the idea that the SIDM proposal can be a promising solution to the TBTF problem.  Figure~\ref{fig:SIDM_kinematics_fit} and figure~\ref{fig:SIDM_TBTF} together  offer an original detailed inspection of the TBTF puzzle.

We eventually utilize the condition $\Gamma \big|_{r_{1}}  \simeq t_{\textrm{age}}^{-1}$ to provide a data-driven estimate of the SIDM cross section. Practically, we need to marginalize over dSph age: We assume $t_{\textrm{age}}$ to be flatly distributed in the range 8--12 Gyr, motivated by the observational indicators pointing MW dSphs to be early-type galaxies \citep{Battaglia13}. 
We collect our estimate of the averaged cross section per unit mass and particle velocity in the last two columns of table~\ref{tab:SIDM_dSph_fit}. 
Notably, while the probed range of averaged velocities turns out to be relatively similar among  MW classicals, spanning $\sim 30$--70 km/s within 68\% h.p.d., the velocity averaged SIDM cross section covers three orders of magnitude. In figure~\ref{fig:SIDM_xsec_histo} we zoom on this aspect directly investigating  the posterior distribution of the DM self-scattering cross section per unit mass, equation~(\ref{eq:SIDM_xsec}). We find Ursa Minor, Draco, Leo~I, and Leo~II  consistently probing cross sections $\sim 0.1$--1.0 cm$^{2}$g$^{-1}$ within 68\% h.p.d.. We also observe Sextans and Fornax to be sensitive to very large cross-section values, with $\sigma /m$ posterior p.d.f. peaked around $\sim 20$ and 40 cm$^{2}$g$^{-1}$ respectively. Sculptor and Carina sit in between the two groups, with the former pointing to $\mathcal{O}(1)$ cm$^{2}$g$^{-1}$, while the latter sharing good overlap with Sextans in $\sigma/m$ posterior p.d.f..

The result depicted in figure~\ref{fig:SIDM_xsec_histo} can be summarized as follows: Within the SIDM model respecting the scatter of the concentration-mass relation read out from the pure CDM simulation presented in \cite{Vogelsberger16}, a \textit{diversity} in the DM particle cross section probed by MW dSphs is strictly demanded by the measured stellar kinematics. Looking at 68\% h.p.d., five of the classical dSphs probe particle cross section well consistent with $\sigma/m \sim 0.5$--3 cm$^{2}$g$^{-1}$, an interval highlighted in \citep{Kaplinghat16,Kamada17} with the analysis of several low-surface brightness and spiral dwarf galaxies within the same SIDM halo model here considered. However, Carina, Sextans and Fornax limit this qualitative interpretation towards a consistent SIDM picture, with Sextans already ruling out $\sigma/m \leq$ 3 cm$^{2}$g$^{-1}$ within 99.7\% h.p.d.. We remark that figure~\ref{fig:SIDM_xsec_histo} provides a cross-section hierarchy that matches the benchmark trends found in previous SIDM-only simulations \citep{Zavala13,Elbert15,Vogelsberger16}, while encoding for the first time the associated observational error for each of the galaxies. We highlight two plausible explanations to the inferred hierarchy in the cross section. First, environmental effects such as tidal stripping due to the MW baryonic disc could be important \citep{Penarrubia10} for these systems, e.g. increasing the scatter of the concentration-mass relation adopted in this work.
 Interestingly, a TBTF puzzle has been found for dwarf galaxies in the field \citep{GarrisonKimmel14,Papastergis15}, where the environmental effects are absent. It would be of great interest to apply the same minimal approach to scrutinize the TBTF problem in field dwarfs and see whether their stellar kinematics also prefer a diverse range of $\sigma/m$. Second, the presence of very light degrees of freedom in the SIDM model may lead to the matter-power suppression that modifies the concentration-mass relation given in equation~(\ref{eq:CDM_mass_conc_rel}), see \cite{Vogelsberger16}, and the extent to which it may work remains to be investigated in detail. Our present study with a minimal approach for SIDM provides a baseline for the assessment of these effects, which we leave for future investigation.

\end{document}